\documentclass{webofc}
\usepackage[varg]{txfonts}   
\usepackage{hyperref}
\usepackage{comment}
\usepackage{caption}
\usepackage{color}
\usepackage{subfigure}
\usepackage{lineno}
\usepackage{wrapfig}
\newcommand*{\pvfinder}{\texttt{PV-Finder}\xspace}
\begin{document}
\title{Advances in developing deep neural networks for finding primary vertices in proton-proton collisions at the LHC}
%
%

\author{\firstname{Simon} \lastname{Akar}\inst{1} \and
        \firstname{Mohamed} \lastname{Elashri}\inst{1} \and
        \firstname{Rocky Bala} \lastname{Garg}\inst{2}\fnsep\thanks{\email{rocky.bala.garg@cern.ch}} \and
        \firstname{Elliott} \lastname{Kauffman}\inst{3} \and
        \firstname{Michael} \lastname{Peters}\inst{1} \and
        \firstname{Henry} \lastname{Schreiner}\inst{3} \and
        \firstname{Michael} \lastname{Sokoloff}\inst{1}\fnsep\thanks{\email{sokoloff@ucmail.uc.edu}} \and
        \firstname{William} \lastname{Tepe}\inst{1} \and
        \firstname{Lauren} \lastname{Tompkins}\inst{2}
}

\institute{University of Cincinnati 
\and
           Stanford University 
\and
           Princeton University
          }

\abstract{%
  We are studying the use of deep neural networks (DNNs) to identify and locate primary vertices (PVs) in proton-proton collisions at the LHC. Earlier work focused on finding primary vertices in simulated LHCb data using a hybrid approach that started with kernel density estimators (KDEs) derived heuristically from the ensemble of charged track parameters and predicted “target histogram” proxies, from which the actual PV positions are extracted. We have recently demonstrated that using a UNet architecture performs indistinguishably from a “flat” convolutional neural network model.
  We have developed an ``end-to-end” tracks-to-hist DNN that predicts target histograms directly from track parameters using simulated LHCb data that provides better performance (a lower false positive rate for the same high efficiency) than the best KDE-to-hists model studied. This DNN also provides better efficiency than the default heuristic algorithm for the same low false positive rate.
  “Quantization” of this model, using FP16 rather than FP32 arithmetic, degrades its performance minimally.
  Reducing the number of UNet channels  degrades performance
  more substantially.
  We have demonstrated that the KDE-to-hists algorithm developed for LHCb data can be adapted to ATLAS and ACTS data using two variations of the UNet architecture. 
  Within ATLAS/ACTS, these algorithms have been validated against the standard vertex finder algorithm.
  Both variations  produce PV-finding efficiencies similar to that of the standard algorithm and
  vertex-vertex separation resolutions that are significantly better.
 
}
\maketitle
\vspace{-0.2in}
\section{Introduction}
\label{intro}
Reconstruction of proton-proton collision points, referred to as primary vertices (PVs), is critical  for physics analyses conducted by all experiments at the Large Hadron Collider (LHC) and for triggering in LHCb. 
The precise identification of the PV locations, and their other characteristics, enables the complete reconstruction of final states under investigation. 
Moreover, it provides crucial information about the collision environment, which is essential for obtaining accurate measurements. 
The task of PV reconstruction poses a significant challenge across the experiments conducted at the LHC.

The LHCb detector has been upgraded for Run 3 of the LHC so that it can process a five-fold increase in its instantaneous luminosity compared to Run 2
and it has removed its hardware-level trigger in favor
of a pure software trigger~\cite{Aaij:2019zbu}. 
The average number of visible PVs detected in the vicinity of the beam crossing area has 
increased from 1.1 to 5.6. 
In contrast, the ATLAS experiment has observed an average of 40-60 simultaneous collisions (known as pile-up, $\mu$) 
during Run 3 in 2023 
and is expected to see 140-200 simultaneous collisions during the coming high-luminosity phase of the LHC. 
These demanding conditions invite development of new PV reconstruction algorithms to address these challenges.

This document presents the implementation and performance of a family of 
machine learning PV reconstruction algorithms known as \pvfinder for both LHCb and ATLAS.
Conceptually, these algorithms compute one-dimensional Kernel Density Estimators (KDEs) that describe 
where charged track trajectories overlap in the vicinity of the
beamline and use these as input feature sets for convolutional
neural networks (CNNs) that predict {\tt target histograms} that are
proxies for the PV positions.
LHCb has traditionally used heuristically computed KDEs with its CNNs;
in this papers it reports merging a fully connected neural network for KDE computation  with 
a CNN to produce an "end-to-end" {\tt tracks-to-hist} deep neural network (DNN) model
and compares its performance with that of older models. 
ATLAS currently uses an analytical approach for KDE computation (referred to as a {\tt KDE-to-hist} model) and compares the performance with the Adaptive Multi-Vertex Finder (AMVF) algorithm~\cite{ATLAS:2019jmx}, 
the heurustic PV identification algorithm currently used in ATLAS. 

\section{\pvfinder in LHCb}
The original LHCb DNN for reconstructing PVs used a
single kernel density estimator (KDE)
calculated using a heuristic algorithm as the input feature set for each event (each beam crossing)
and produced a target histogram from which PV positions were deduced.
We refer to this class of algorithms as {\tt KDE-to-hist} algorithms.
The results of the initial proof-of-principal project, and some details of the
``toy Monte Carlo" and KDE used for that study are reported
in Ref.~\cite{Fang:2019wsd}.
Using track parameters produced by the LHCb Run~3 Vertex Locator (VELO) tracking algorithm~\cite{Hennequin:2019itm} leads to significantly better performance~\cite{akar2020updated}.
Since then, our research has advanced in several directions.
We replaced our original input feature set with four input feature sets:
a first KDE based on summed probabilities in voxels projected onto the beam axis,
a  second KDE based on summed (probability-squared values) in voxels projected onto the beam axis,
plus the $ x- $ and $ y- $ coordinates of the maximum summed probability at each
value of $ z $ (along the beam axis).
We found that using a modified U-Net architecture~\cite{ronneberger2015unet}
in place of our
original CNN architecture provided equally good
fidelity and trained much more quickly.
We also investigated using a fully connected network to calculate
a KDE from track parameters (a {\tt tracks-to-KDE} model)
and merging this model with a {\tt KDE-to-hist} model to
produce an ``end-to-end" {\tt tracks-to-hist} neural network.
The fidelity of the {\tt tracks-to-hist} studied then
was inferior to that of the {\tt KDE-to-hist} models.
The results of these studies were presented at CHEP-2021~\cite{Akar:2021gns}.

A major advance reported at this conference (CHEP-2023) is that we
have produced a {\tt tracks-to-hist} model that produces efficiencies
very similar to the best produced by our {\tt KDE-to-hist} models
{\em and} produces significantly lower false positive (FP) rates.
These results were reported previously at ACAT-2022~\cite{Akar:2023zhd}.
Below, we summarize the most salient features.
Brand new for this conference are results using FP16 arithmetic rather
than FP32 arithmetic for the {\tt tracks-to-hist} model
and results using smaller U-Net components in the 
FP16 {\tt tracks-to-hist} models.

\begin{figure}[t]
\centering
\includegraphics[width=0.95\textwidth,clip]{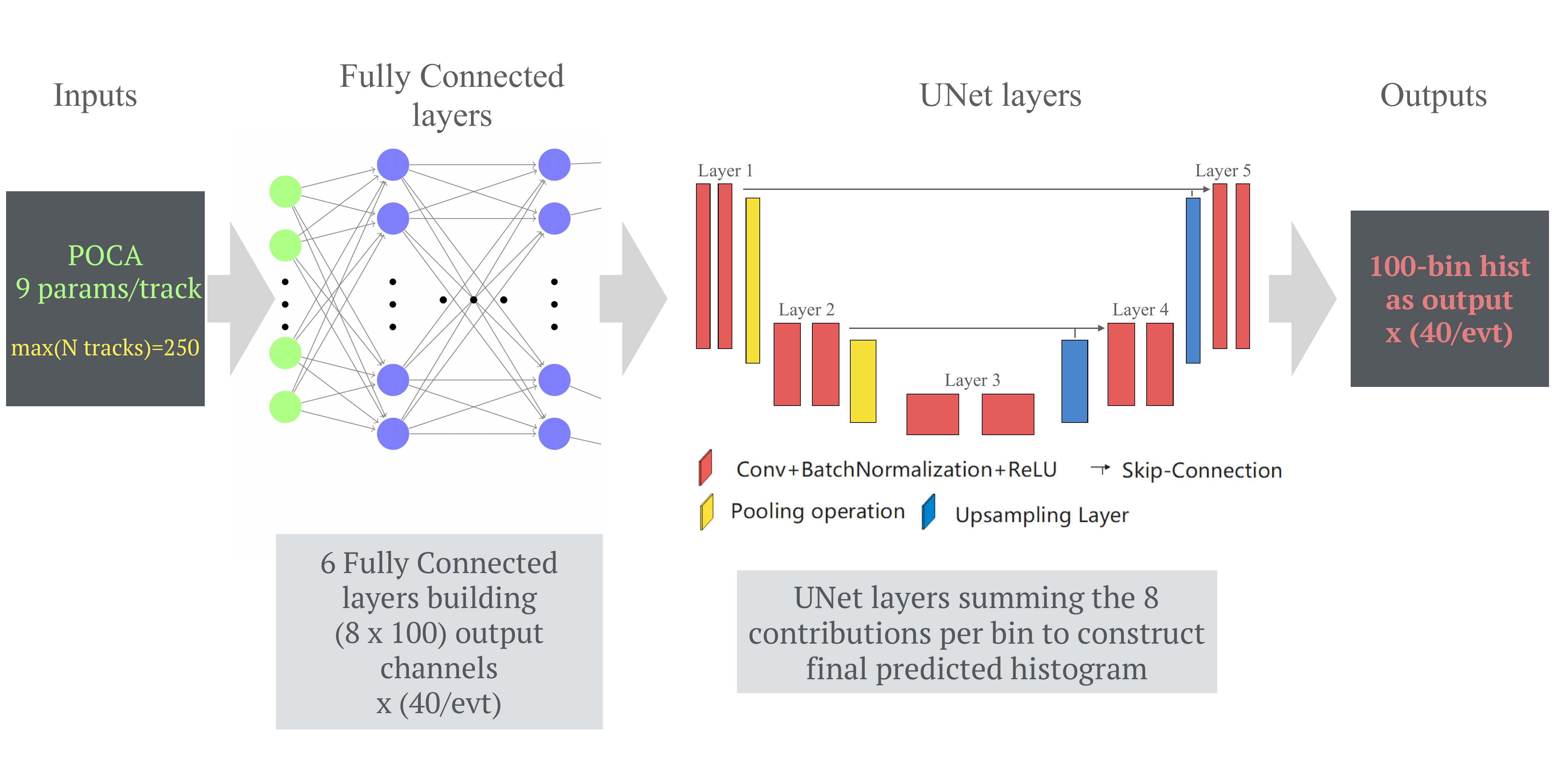}
\vspace{-5pt}
\caption{
This diagram illustrates the end-to-end, {\tt tracks-to-hist}, DNN 
Each event is now sliced into 40 independent 100-bin intervals.
Six fully connected layers populate 8 100-bin channels in sixth layer, for each track.
These contributions are summed and processed by a U-Net model with 5 convolutional layers to construct the final 100-bin histogram.
}
\label{fig:model}
\vskip -0.4in
\end{figure}

The current {\tt tracks-to-hist} model, whose architecture is shown in Fig.~\ref{fig:model}, includes a few updates
relative to the original version described in Ref.~\cite{Akar:2021gns}:
the {\tt tracks-to-KDE} part of the model consists of 6 fully connected layers that are initially trained to
produce a KDE and the weights of the first 5 layers are temporarily
frozen; a variation with 8 latent feature sets is merged to a {\tt KDE-to-hist}-like
DNN where the classical CNN layers are
replaced by a U-Net model.
Critically, we also updated the structure of the input data for training
and inference.
 In the earlier approach~\cite{Akar:2021gns}, 
 the target histograms consisted of 4000 bins along the z-direction (beamline), each $100\,{\rm \mu m}$ wide, spanning the active area of the VELO around the interaction point, such that $z \in [-100, 300]\,{\rm mm}$.
 Parameters describing all tracks served as input features.
In place of describing the true PVs using a single 4000-bin histogram,
we now
slice each event into 40 intervals of 100 bins each.
For each interval, parameters of tracks whose points of closest approach 
to the beamline lie within 2.5 mm of the interval edges are used as input features.
This approach is motivated by the fact that the shapes of the target histogram
are expected to be invariant as a function of the true PV position
and it is easier for a DNN to learn to predict target histograms
over a smaller range of bins.
In particular, the fully connected layers that calculate the
KDE-like latent features used as input features by the U-Net layers 
predict heuristic KDEs as the ground truth much more effectively
when training 100-bin intervals rather than the full 4000-bin range.
Additionally, the depth of the U-Net part of the DNN can be lower
when processing a 100-bin feature set rather than a 4000-bin feature set.
With an average of $ \sim 5 $ PVs per event, most of the bins in both the KDE and target histograms have no significant activity.
We expect this will allow us to eventually build a more performant inference 
engine in the LHCb software stack.
The 40 intervals of 100 bins are independent and homogeneous between events.
Each interval is treated independently, after which the predicted 4000-bin
histogram is stitched back together.
As in past studies, an asymmetry parameter between the cost of overestimating contributions to the
target histograms and underestimating them~\cite{Fang:2019wsd} is
used as a hyperparameter to allow higher efficiency by incurring higher false
positive rates.

\begin{figure}
\centering
\includegraphics[width=0.45\textwidth,clip]{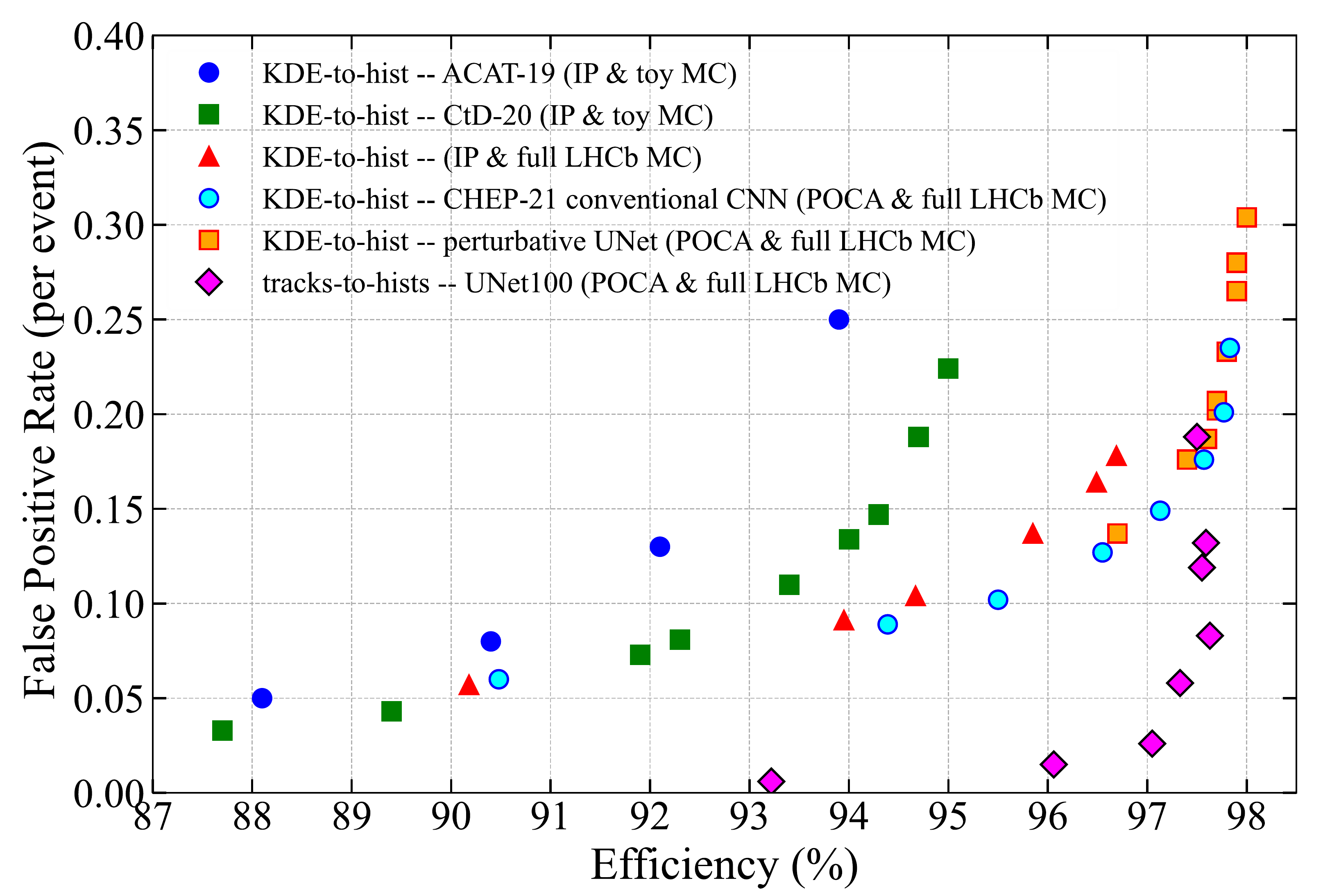}
\includegraphics[width=0.50\textwidth,clip]{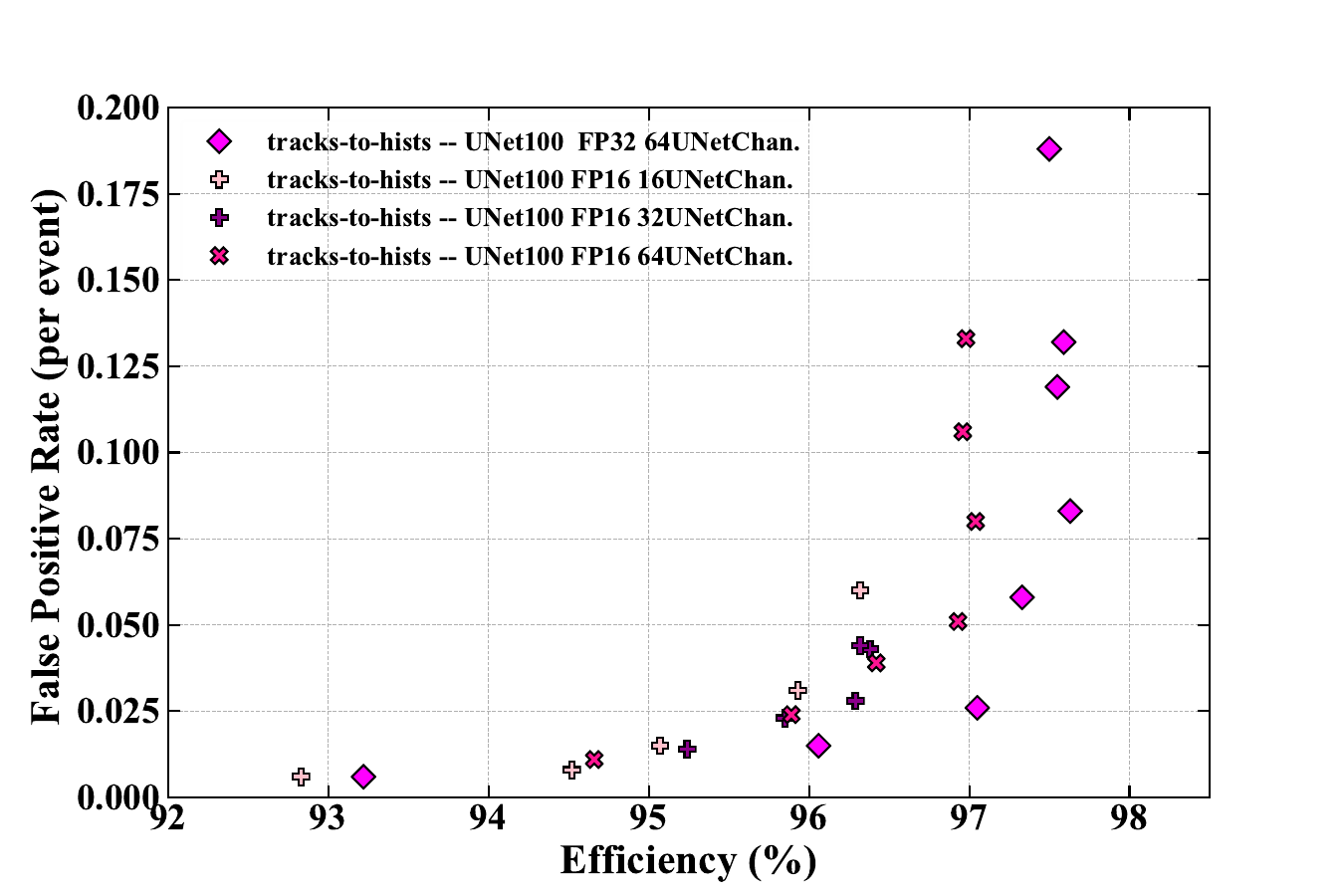}
\vspace{-5pt}
\caption{
(left)
Comparison between the performances of models reported in previous years and the new {\tt tracks-to-hist} model (magenta diamonds). A cost asymmetry parameter described in Ref.~\cite{Fang:2019wsd}
is varied to produce the families of points observed.
(right)
Comparison between {\tt tracks-to-hist} models. The magenta diamonds here are the same
as in the plot on the left.
The other models have U-Net architectures but use FP16 arithmetic
rather than FP32.
Two of the FP16 models have smaller U-Net components than the FP32 model.
NB: the horizontal and vertical scales on the right cover
more limited ranges than those on the left.
}
\label{fig:evolution}
\vskip -0.3in
\end{figure}

Performance is evaluated using a heuristic algorithm, based on the PV positions along the beam axis, $z$.
Exactly how efficiencies and FP rates are calculated is described in
Ref.~\cite{Akar:2021gns}.
The left-hand plot in Fig.~\ref{fig:evolution} shows how the performance of the 
DNN algorithms have evolved over time.
The efficiency is shown on the horizontal axis and the false positive rate per event is shown on the vertical axis.
The solid blue circles show the performance of any early {\tt KDE-to-hist} model described at ACAT-2019~\cite{Fang:2019wsd}.
The green squares show the performances of a {\tt KDE-to-hist} described at Connecting-the-Dots in 2020~\cite{akar2020updated}.
Both of the above models were trained using ``toy Monte Carlo” with proto-tracking.
All subsequent DNN models were trained using full VELO tracking algorithm~\cite{Hennequin:2019itm}, leading to significantly better performances (red triangles to be compared to green squares).
The cyan circles and the yellow squares correspond to the best achieved performances for {\tt KDE-to-hist} models using either a classical CNN architecture or the U-Net model described at CHEP-2021

The performances of all above models were obtained using an ``older" matching procedure with a fixed search window of $0.5\,{\rm mm}$.
The magenta diamonds show the performance of the {\tt tracks-to-hist} model described above
using the matching procedure described in
Ref.~\cite{Akar:2021gns}.
The  new {\tt tracks-to-hist} model enables the DNN
to simultaneously reach high
efficiencies ($>97\%$) and low false positive rates ($0.03$ per event or $0.6\%$
per reconstructed PV).

Running an inference engine inside a software stack adds another ``knob to turn" --
throughput versus fidelity.
Computing resources are finite, especially in LHCb's first level software trigger
which processes 30 MHz of beam crossing data, about 40 Tbit/s, in a GPU application~\cite{Aaij:2019zbu}.
Modern GPUs provide FP16 performance that can be about twice as
fast as FP32 arithmetic, so it is interesting to investigate whether using
FP16 arithmetic degrades performance significantly.
It similarly interesting to investigate how performance degrades
as the size of the convolutional network inside our DNN is reduced.
The right-hand plot in Fig.~\ref{fig:evolution} shows the efficiency versus
FP rate for four DNN configurations.
The magenta diamonds correspond to the default {\tt tracks-to-hist} configuration.
These points are exactly the same as those in the left-hand plot;
the ranges of the axes have been modified to focus on the region of interest.
The purple "$ \times $" markers correspond to the same logical configuration,
but using FP16 arithmetic rather than FP32. 
Near 96\% efficiency, the FP rate has increased marginally.
Near 97\% efficiency, the FP rate has increased much more substantially.
Reducing the number of U-Net channels from 64 to 32 or 16,
while using FP16 arithmetic,
(the darker and lighter crosses in the plot)
additionally additionally increases the FP rate near 96\% efficiency by a small
amount, but increases the FP rate much more significantly near 96.5\%.
We have begun to code an inference engine to run in LHCb's first level
software trigger.
The details of the model to be instantiated will balance
fidelity of the model against throughput.

\vspace{-0.1in}
\section{\pvfinder in ATLAS}
The ATLAS experiment at the LHC is a versatile particle detector designed with a symmetric cylindrical geometry and near-complete coverage of 4$\pi$ in solid angle~\cite{atlascollaboration2023atlas}. It has a multi-layer structure with many sub-detector systems including an inner tracking detector, superconducting magnets, electromagnetic and hadronic calorimeters, and a muon spectrometer. An extensive software suite~\cite{ATL-SOFT-PUB-2021-001} facilitates its various functions such as data reconstruction and analysis, detector operations, trigger and data acquisition systems etc. 

\begin{figure}[!h]
\begin{center}
    \includegraphics[width=0.90\textwidth]{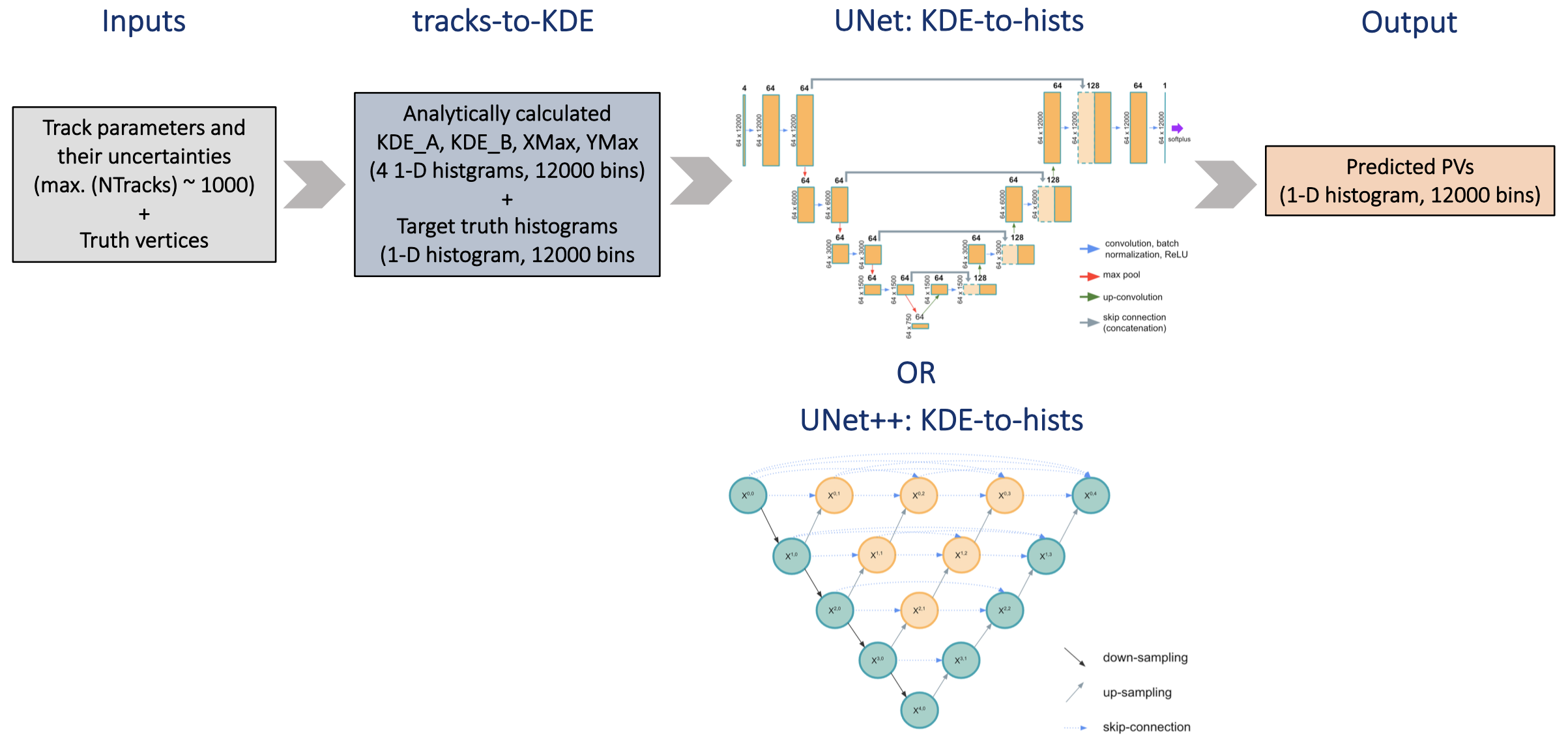}
    \caption{Flowchart representing work-flow of the \pvfinder algorithm from left to right.}
    \label{fig:model_atlas}
    \end{center}
\end{figure}

The input dataset used for studying \pvfinder in ATLAS has been generated using 
{POWHEG BOX}[v2]~\cite{Alioli:2010xd} interfaced with PYTHIA[8.230]~\cite{Sjostrand:2014zea} 
and processed through the ATLAS detector simulation framework~\cite{ATL-SOFT-PUB-2021-001}, 
using the GEANT4 toolkit~\cite{GEANT4:2002zbu}. 
The hard-scatter (HS) process involves the production of semi-leptonically decaying top quark pairs ($t\bar{t}$) from proton-proton collisions at a center-of-mass energy of 13~TeV,  
overlaid with simulated minimum-bias events with an average pile-up of 60.
\vskip 0.2in
\leftline{\bf 3.1 \hspace{0.05in} \pvfinder algorithm and model architecture}
\vskip 0.1in\noindent
The flowchart representing work-flow of the \pvfinder algorithm for ATLAS is shown in figure~\ref{fig:model_atlas}. More details about the architecture can be found at the ATLAS PubNote~\cite{ATLAS_pvfinder_pubnote}. 
Truth-matched reconstructed tracks passing tight quality selection cuts~\cite{ATL-PHYS-PUB-2015-051} and p$_{\textrm{T}}$ $>$ {500~MeV} are used for the preparation of input features for the neural network. A track's signed radial and longitudinal impact parameters, 
$d_{0}$ and $z_{0}$,  measured at the point of closest approach (POCA) to the beamline, and their uncertainties, $\sigma(d_{0})$ and
$\sigma(z_{0})$, are used as input to generate KDEs. Each KDE feature is a one-dimensional binned histogram with 12,000 bins in $z \in [-240, 240]\,{\rm mm}$, corresponding to a bin-size of 40~$\mu$m.

\vspace{-0.3in}

To compute these features, each track is modeled as a correlated radial and longitudinal Gaussian probability distribution $\mathbb{P}(d,z)$ centred at ($d_{0}, z_{0}$) which is defined as follows:
\begin{equation}
    \mathbb{P}(r) = \mathbb{P}(d,z)= \frac{1}{2\pi\sqrt{|\Sigma}|} \text{exp}\bigg(-\frac{1}{2}\Big((d - d_0),(z - z_0)\Big)^{T}\Sigma^{-1}\Big((d - d_0),(z - z_0)\Big)\bigg)
\end{equation}
where $d$ and $z$ are coordinates in the radial and longitudinal directions and $\Sigma$ = $\left( {\begin{array}{cc}
        \sigma^{2}(d_{0}) & \sigma(d_{0}, z_{0}) \\
            \sigma(d_{0}, z_{0}) &  \sigma^{2}(z_{0})
  \end{array} } \right)$ is the covariance matrix. The sum of probabilities from all the contributing tracks is considered in each $z$-bin and four KDE features are constructed: \texttt{KDE-A} (sum of track probability values), \texttt{KDE-B} (sum of the squares of track probability values), \texttt{XMax} (\texttt{YMax}) (location of the maximum summed track probability in $x$($y$) (mm)).
 An example illustrating these four features for a random event is shown in Fig.~\ref{fig:KDE_features}.
 The vertical grey lines in the upper plot mark the locations of true primary vertices while horizontal grey line in the lower plot denotes  the position of the beam spot in the radial direction. A restricted range of the luminous region is shown so that details can be seen.

 To train the neural network, a one-dimensional target truth histogram, with the same binning as the input features and calculated by considering Gaussian probabilities around truth vertex locations, is also provided as input along with the four KDE features. A CNN is trained on these features which then outputs a distribution with approximately Gaussian peaks centered at the predicted locations of PVs.
An algorithm then takes this predicted distribution and identifies the candidate PV locations on the $z$-axis by finding the local maxima. Two NN architectures have been considered for these studies: the UNet architecture is inspired from the original architecture developed for biomedical image segmentation~\cite{ronneberger2015unet} while the UNet++ architecture is a variation of UNet with dense skip connections.

\begin{minipage}{\textwidth}
   \begin{minipage}[t]{0.45\textwidth}
    \centering
    \includegraphics[width=0.95\textwidth]{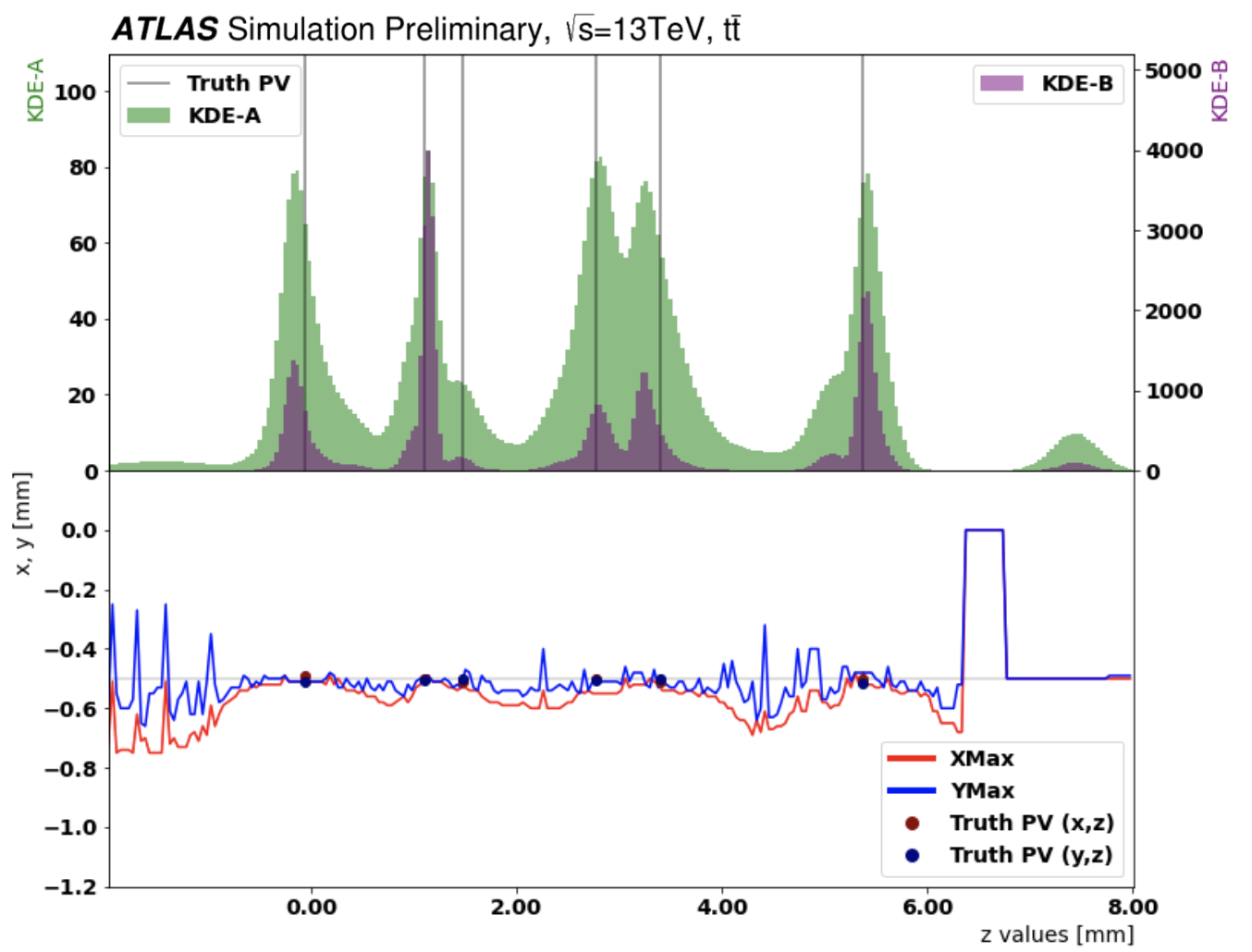} 
    \captionof{figure}{An example of the input KDE features. Upper plot shows \texttt{KDE-A} (green) and \texttt{KDE-B} (purple) while lower plot shows \texttt{XMax} (blue) and \texttt{YMax} (red). See text for details.}
    \label{fig:KDE_features}
  \end{minipage}
  \hspace{0.1in}
  \begin{minipage}[t]{0.45\textwidth}
    \centering
    \includegraphics[width=0.95\textwidth]{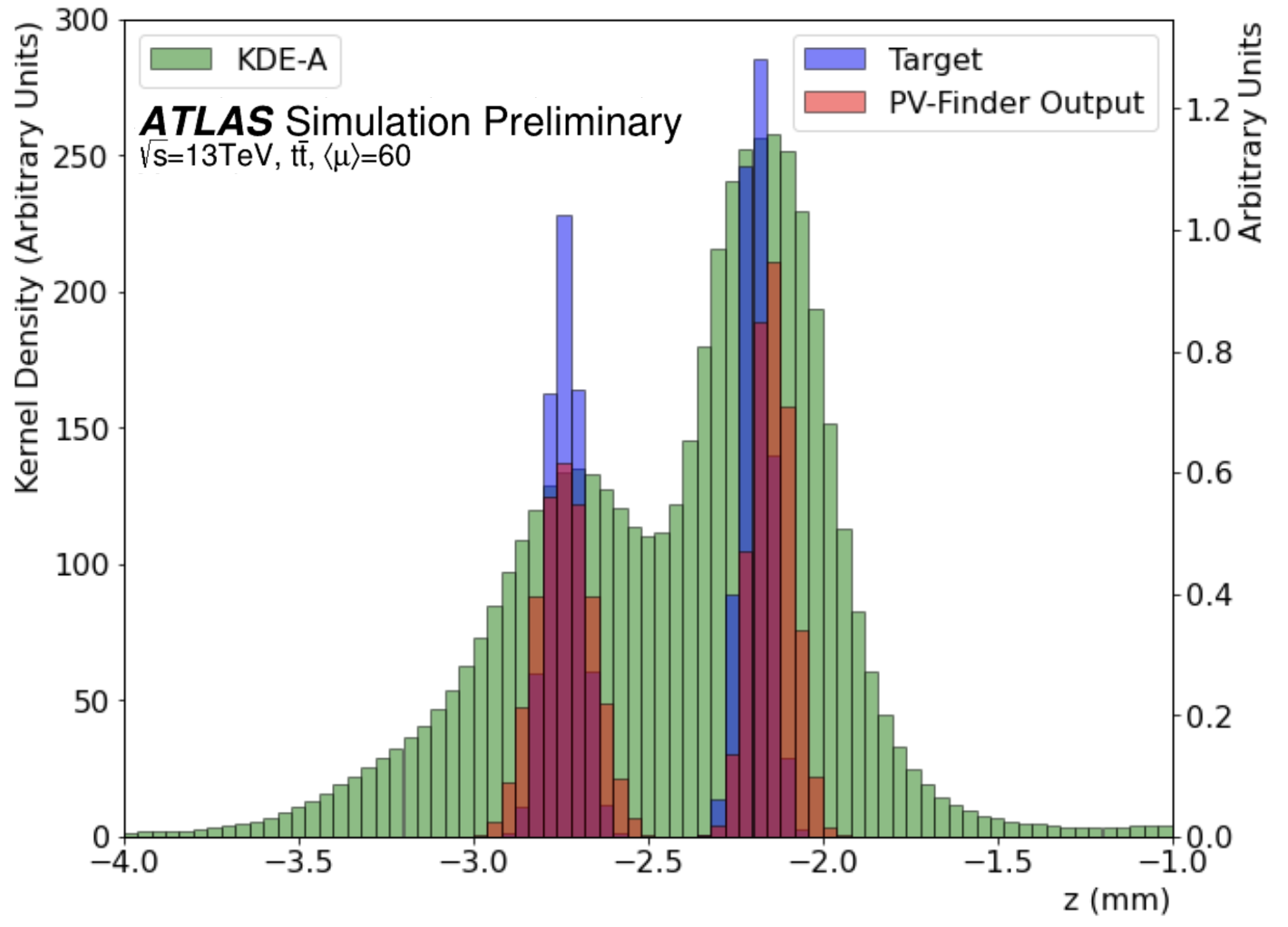} 
    \captionof{figure}{An example of correctly predicted \pvfinder output (in red) compared with KDE-A (in green) and target truth (in blue).}
    \label{fig:KDE_output}
  \end{minipage}
  \hspace{0.1in}
\end{minipage}
\vskip 0.1in
\leftline{\bf 3.2 \hspace{0.05in} Performance}
The \pvfinder algorithm's performance for UNet and UNet++ architectures has been studied and a comparative analysis is conducted with the AMVF algorithm using an independent test data sample. Figure~\ref{fig:KDE_output} showcases an example of two adjacent vertices accurately located by the \pvfinder algorithm.
To quantitatively evaluate the performance of the \pvfinder, 
vertex classification is performed, and
efficiency and false positive rates are calculated. 
The classification assigns vertices into distinct categories, namely clean, merged, split, and fake based on the distance between the center of a predicted vertex and the $z$-location of truth vertices. The classification is illustrated in Figure~\ref{fig:VtxScheme} and demonstrated in Figure~\ref{fig:VtxClassification} for the three approaches.

The truth and reconstructed primary vertices are associated based on a vertex-vertex resolution, $\sigma_{\textrm{vtx-vtx}}$, 
which is obtained by computing the $z$-difference between pairs of nearby reconstructed vertices and fitting the distribution with the fit function: $y$ = $\frac{a}{1+\exp{(b\cdot(R_{cc} - |x|))}}+c$, where $a, b, c$ are free parameters, and $R_{cc}$
is the cluster–cluster resolution referred to as $\sigma_{\textrm{vtx-vtx}}$. The vertex-vertex resolution for \pvfinder UNet, \pvfinder UNet++ and AMVF is presented in Figure~\ref{fig:ResZ} and Table~\ref{tab:results}.

The vertex finding efficiency is defined as the number of truth vertices assigned to reconstructed vertices as ``clean'' and ``merged'' divided by the total number of reconstructable truth vertices while the false positive rate is defined as the average number of predicted vertices not matched to any truth vertex. Figure~\ref{fig:Eff_vs_ntrks} shows the vertex finding
efficiency as a function of the number of reconstructed tracks associated to a truth vertex and Table~\ref{tab:results} shows the average efficiency and false positive rates obtained for three cases. 

\begin{minipage}{\textwidth}
  \begin{minipage}[t]{0.40\textwidth}
    \centering
    \includegraphics[width=0.60\textwidth]{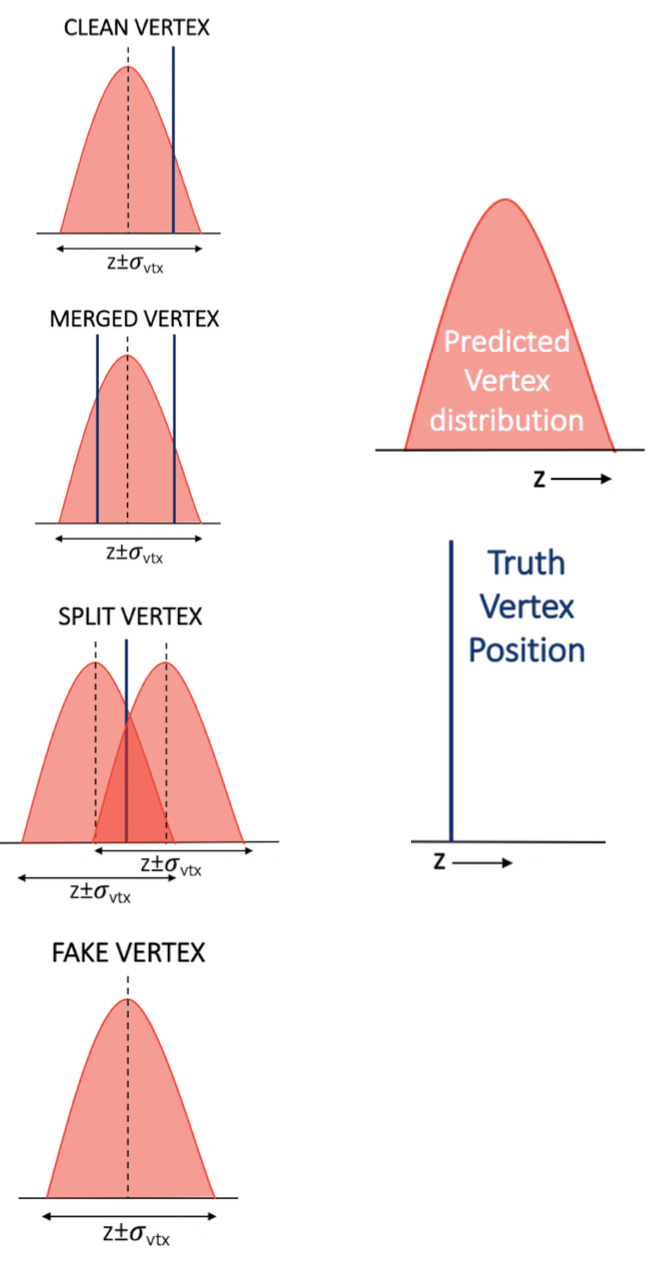} 
    \captionof{figure}{Vertex Classification Scheme.  The dashed line represents the reconstructed vertex location, the solid line represents the truth vertex location, and $\sigma_{\textrm{vtx}}$ represents $\sigma_{\textrm{vtx-vtx}}$.}
    \label{fig:VtxScheme}
  \end{minipage}
  \hspace{0.2in}
   \begin{minipage}[t]{0.5\textwidth}
    \centering
    \includegraphics[width=0.95\textwidth]{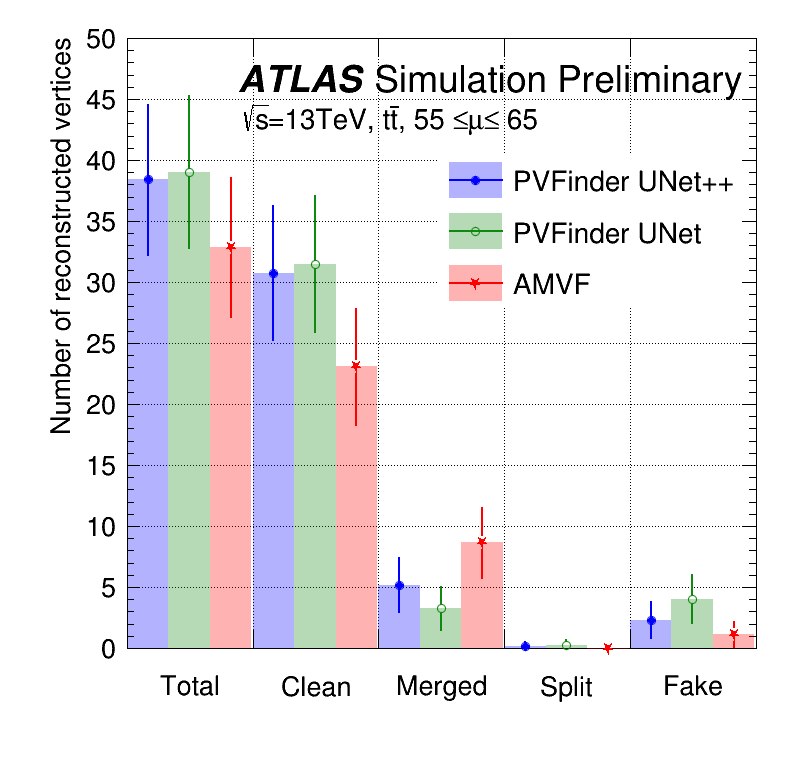} 
    \captionof{figure}{Vertex Classification rates for \pvfinder and AMVF for a $t\bar{t}$ sample with $55$ $\leq$ $\mu$ $\leq$ $65$, comparing AMVF (red), \pvfinder UNet (green) and \pvfinder UNet++ (blue).}
    \label{fig:VtxClassification}
  \end{minipage}
  \hspace{0.1in}
\end{minipage}
\vspace{0.2in}
\hspace{-0.2in}
\begin{minipage}{\textwidth}
  \begin{minipage}[t]{0.5\textwidth}
    \centering
    \includegraphics[width=0.95\textwidth]{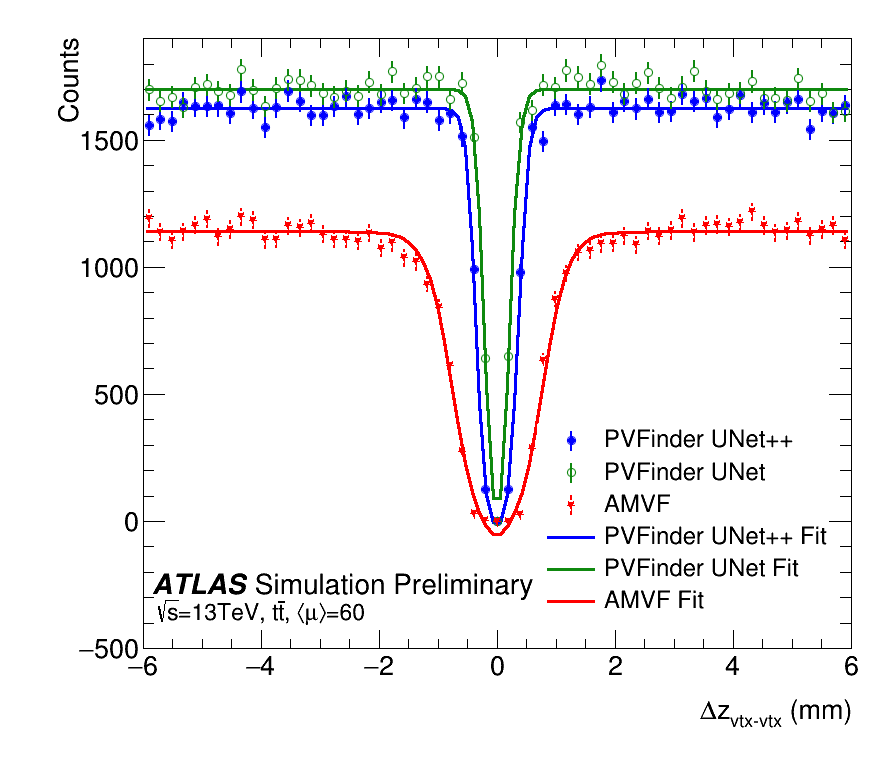} 
    \captionof{figure}{${\Delta}z_{\textrm{vtx-vtx}}$ for \pvfinder UNet++ (blue), \pvfinder UNet (green) and AMVF (red). See text for discussion. More details at~\cite{ATLAS_pvfinder_pubnote}.
    }
    \label{fig:ResZ}
  \end{minipage}
  \hspace{0.1in}
   \begin{minipage}[t]{0.5\textwidth}
    \centering
    \includegraphics[width=0.95\textwidth]{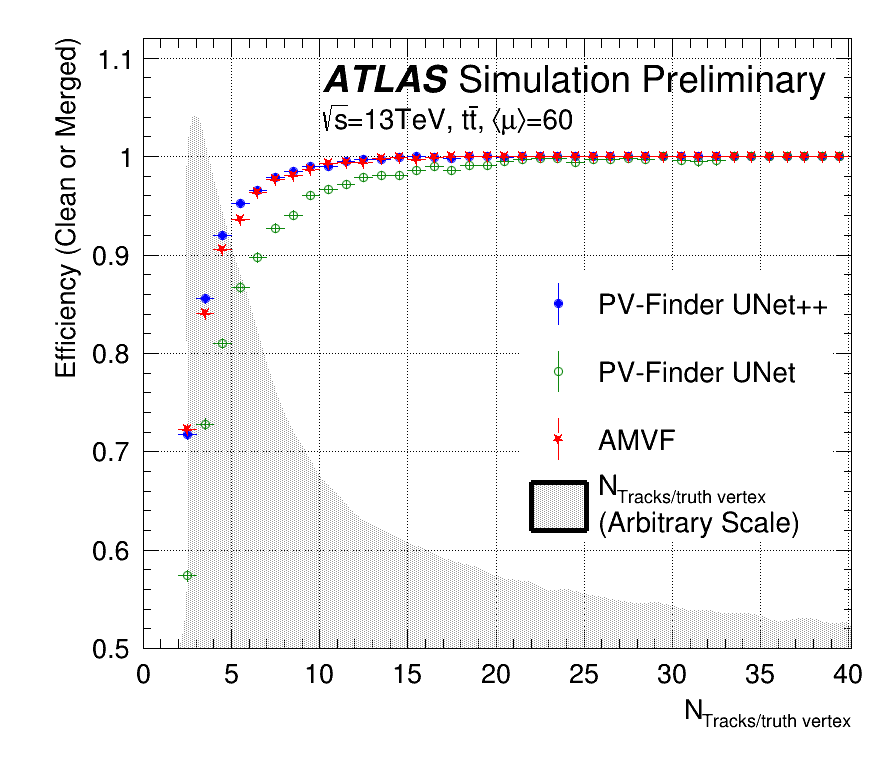} 
    \captionof{figure}{PV efficiencies for AMVF and  \pvfinder's UNet++ and UNet architectures vs the number of reconstructed tracks associated to truth vertex. More details at~\cite{ATLAS_pvfinder_pubnote}.}
    \label{fig:Eff_vs_ntrks}
  \end{minipage}
  \hspace{0.1in}
\end{minipage}
\vspace{-0.2in}
\begin{table}[h]
\centering
    \begin{tabular}{llcc}
    \hline
    Method    & $\sigma_{\textrm{vtx-vtx}}$ (mm) & Efficiency & FP Rate (per event) \\ \hline
    \pvfinder UNet & $0.23 \pm 0.01$  & 88.7\%     & 2.6\\
    \pvfinder UNet++ & $0.37 \pm 0.01$ & 94.2\% & 1.5\\
    AMVF & $0.76 \pm 0.02$  & 93.9\% & 0.8\\ \hline
    \end{tabular}
    \caption{A comparison of the vertex-vertex separation resolutions, the efficiencies,
    and the false positive (FP) rates of the algorithms.}
    \label{tab:results}
  \vskip -0.2in
\end{table}
\newpage
\section{Conclusion}
The \pvfinder family of algorithms has been studied by both the LHCb and ATLAS experiments. 
LHCb has demonstrated the performances of the end-to-end {\tt tracks-to-hist} approach for several
configurations including those that use FP16 arithmetic rather than FP32. 
ATLAS has demonstrated that a hybrid {\tt KDE-to-hist} approach produces
efficiencies comparable to the ATLAS AMVF algorithm while also achieving significanty improved resolution. These enhanced efficiency and resolution metrics hold significant importance, especially considering the future High Luminosity LHC program. 
The results are promising and motivate further studies and refinement of the \pvfinder algorithms across experiments.
\\\\
{[Copyright 2023 CERN for the benefit of the ATLAS and LHCb Collaborations. CC-BY-4.0 license]}
\\\\
{[This work was supported by the National Science Foundation under Cooperative Agreements OAC-1836650 and PHY-2323298]}
\bibliography{CHEP2023_PVFinder.bib}

\end{document}